\documentclass[fleqn,twoside]{article}
\usepackage{amsmath}
\usepackage{espcrc2}
\usepackage{graphicx}

\title{Deconfinement and Chiral Restoration in Hot and Dense Matter}
\author{Kenji Fukushima
\thanks{We acknoledge the support of the Japan Society for the
Promotion of Science for Young Scientists and the U.S.\ Department of
Energy under cooperative research agreement \#DF-FC02-94ER40818.}
\address{Center for Theoretical Physics, MIT,
         77 Mass.\ Ave., Cambridge, MA 02139, USA}
\address{Department of Physics, University of Tokyo,
         7-3-1 Hongo, Bunkyo-ku, Tokyo 113-0033, Japan}}

\begin{document}

\begin{abstract}
We propose a picture that the chiral phase transition at zero quark
mass and the deconfinement transition at infinite quark mass are
continuously connected. This gives a simple interpretation on the
coincidence of the pseudo-critical temperatures observed in lattice
QCD. We discuss a possible dynamical mechanism behind the simultaneous
crossovers and show the results in a model study.
\end{abstract}

\maketitle

It is widely accepted that Quantum Chromodynamics (QCD) has phase
transitions at high temperature, the nature of which depends on the
quark mass, $m_{\rm q}$. In the limit of $m_{\rm q}=0$, QCD has chiral
symmetry, $\mathrm{SU_L}(N_{\rm f})\times\mathrm{SU_R}(N_{\rm f})$, in
the case of $N_{\rm f}$ flavors, that spontaneously breaks into
$\mathrm{SU_V}(N_{\rm f})$ at low temperatures. The chiral condensate,
$\langle\bar{q}q\rangle$, serves as an order parameter for chiral
restoration at high temperature. Since the quark mass term breaks
chiral symmetry, $m_{\rm q}/f_\pi$ is regarded as the magnitude of the
explicit chiral symmetry breaking. When $m_{\rm q}=\infty$, on the
other hand, there is no dynamical quark and thus no remnant of chiral
symmetry at all. QCD is reduced to a pure gluonic theory, which has
center symmetry at finite temperature, that is $\mathrm{Z}(N_{\rm c})$
for $N_{\rm c}$ colors. The order parameter is given by the Polyakov
loop, $\langle l\rangle=\langle\text{tr}L\rangle=\langle\text{tr}
\mathcal{T}\mathrm{e}^{-\mathrm{i}\int\!\mathrm{d}x_4 A_4}\rangle$.
Center symmetry is broken by thermal quark excitation
and the magnitude of the explicit center symmetry breaking is
$\mathrm{e}^{-M_{\rm q}/T}$ where $M_{\rm q}$ is the constituent quark
mass.

In the limits of $m_{\rm q}=0$ and $m_{\rm q}=\infty$ the nature of
the QCD phase transition can be predicted from generic arguments based
on the universality class. It has been almost established today that
the chiral phase transition at $m_{\rm q}=0$ is of second order for
$N_{\rm f}=2$ and first order for $N_{\rm f}\geq 3$~\cite{pis84}, and
that the deconfinement phase transition at $m_{\rm q}=\infty$ is of
second order for $N_{\rm c}=2$ and first order for
$N_{\rm c}\geq 3$~\cite{sve82}. Thus, a question arises; what is the
QCD phase transition for $0<m_{\rm q}<\infty$?

\textit{A naive expectation} would be the following. The deconfinement
and chiral phase transitions are different phenomena lying in the
opposite limits. The critical temperature is known from lattice QCD to
be $T_\chi\simeq 150\,\text{MeV}$ for chiral restoration and
$T_{\rm d}\simeq 270\,\text{MeV}$ for deconfinement. In the presence
of finite $m_{\rm q}$ they are both blurred by the explicit symmetry
breaking. One would, as a result, expect to find a crossover
associated with chiral restoration near $T_\chi$ and another crossover
associated with deconfinement near $T_{\rm d}$.

\textit{The fact} turned out different. In the lattice QCD simulation
the chiral condensate and the Polyakov loop have been measured as a
function of the temperature. Contrary to the naive expectation, only
one crossover has been found. The chiral condensate and the Polyakov
loop indicate a crossover simultaneously at the same
temperature, $T_{\rm c}$, and moreover, the chiral susceptibility has
one peak at $T=T_{\rm c}$, and so does the Polyakov loop
susceptibility~\cite{fuk86}.

\textit{A naive explanation} would be the following. Since there is
only one crossover, there must be either the chiral or deconfinement
phase transition in reality. A peak in the susceptibility signifies a
remnant of the second-order phase transition in which the
susceptibility has a singularity at the critical point. One might
think that only the chiral phase transition can be regarded as an
approximate second-order phase transition, so that the susceptibility
peak comes only from the chiral phase transition and no remnant of the
deconfinement transition remains. This explanation is not correct,
however.

Even though the above explanation could work in the vicinity of the
chiral limit, the lattice data is taken for various quark masses
corresponding to the pion mass ranging from $\sim400\,\text{MeV}$ to
several GeV, that are not so close to the chiral limit. It
could be even possible that the lattice data is closer to the
deconfinement transition rather than the chiral phase transition. The
point is that there is a second-order phase transition not only near
the chiral limit but also near the heavy quark mass limit. In the case
of $N_{\rm f}=3$ and $N_{\rm c}=3$, as mentioned before, both the
chiral and deconfinement phase transitions in the limit of
$m_{\rm q}=0$ and $m_{\rm q}=\infty$ are of first order. The phase
transition is smeared by the effect of the explicit symmetry breaking
and eventually becomes a crossover. The point where the phase
transition ceases to be of first order is called the critical
end-point (CEP), at which the system undergoes a second-order phase
transition. Interestingly enough, the lattice and model studies give
the quark mass around $800\,\text{MeV}$ at the deconfinement CEP, that
seems not to be very heavy~\cite{att88,fuk03,fuk04}.

\begin{figure}
\begin{center}
\includegraphics[width=5cm]{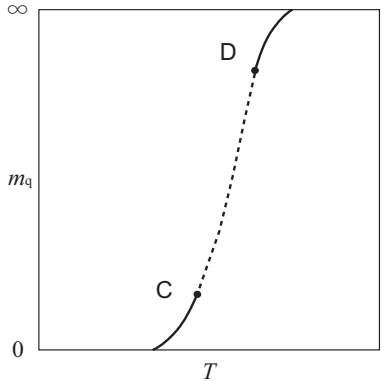}
\vspace{-5mm}
\caption{Schematic picture of the QCD phase transition taken from
Ref.~\cite{hat04}.}
\label{fig:phase}
\end{center}
\end{figure}

\textit{A correct interpretation} on the lattice data should be the
following. One must consider both the chiral and deconfinement
transitions on an equal footing. When $m_{\rm q}$ is small, there is a
second-order phase transition at the chiral CEP, denoted by
\textsf{C}, and a second-order phase transition at the deconfinement
CEP, denoted by \textsf{D} in Fig.~\ref{fig:phase}. The susceptibility
peak may reflect $\textsf{C}$ or $\textsf{D}$ or possibly both for a
quark mass of order hundreds MeV. The fact is, however, there appears
only one peak in the susceptibility for all $m_{\rm q}$. This means
that \textsf{C} and \textsf{D} are connected smoothly by a single
crossover boundary as shown by the dashed curve in
Fig.~\ref{fig:phase}~\cite{hat04}. We would emphasize that this is the
only one possible interpretation based on the fact observed on the
lattice.

The nature of the QCD phase transition is highly non-trivial rather
than what has been argued naively. There is one phase transition
that is a mixture of chiral restoration and deconfinement, neither of
which loses its physical meaning. Then, a question one will come
across next would be how two distinct phenomena become locked together
\textit{dynamically}. Since no generic argument is applicable due to
the explicit breaking of chiral and center symmetries, it should be a
dynamical problem depending on the coupling between the chiral
condensate (or the $\sigma$ meson) and the Polyakov loop.

The coupling proposed in Ref.~\cite{fuk04} is given by
\begin{equation}
 \begin{split}
 &\sim\int\frac{\mathrm{d}^3 p}{(2\pi)^3}\mathrm{Tr_c}\biggl\{\ln
  \bigl[1+L\mathrm{e}^{-(E_p-\mu)/T}\bigr] \\
 &\qquad\qquad\qquad\quad +\ln\bigl[1+L^\dagger
  \mathrm{e}^{-(E_p+\mu)/T}\bigr]\biggr\},
 \end{split}
\label{eq:coupling}
\end{equation}
where $\mathrm{Tr_c}$ is the trace over color and $\mu$ a quark
chemical potential. The quasi-quark energy is
$E_p=\sqrt{p^2+M_{\rm q}^2}$ where $M_{\rm q}^2$ depends on the chiral
condensate, which gives the coupling between the chiral condensate and
the Polyakov loop.

\begin{figure}
\begin{center}
\includegraphics[width=5cm]{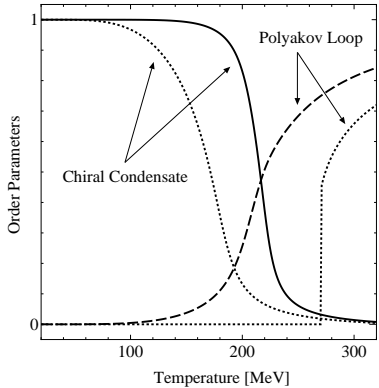}
\vspace{-5mm}
\caption{The chiral condensate and the Polyakov loop as a function of
the temperature.}
\label{fig:order}
\end{center}
\end{figure}

Fig.~\ref{fig:order} shows the results from a model for the chiral
condensate and the Polyakov loop with the coupling
(\ref{eq:coupling}). The deconfinement transition is, as shown by the
dotted curve, of first order without dynamical quarks. The chiral
condensate shown by another dotted curve is the result of the
standard NJL model with $u$ and $d$ quarks having the mass,
$m_{\rm q}=5.5\,\text{MeV}$. They are shown for reference. Obviously
the chiral crossover (solid curve) and the deconfinement crossover
(dashed curve) occur around the same pseudo-critical temperature as a
result of the coupling (\ref{eq:coupling}).

The reason two crossovers come closer to each other is understood from
the generic property of the coupling (\ref{eq:coupling}). Roughly
speaking, if the expectation value of the Polyakov loop is small, the
quark excitation is suppressed because $L$ is in front of the thermal
factor of the quark excitation. This means that chiral restoration
cannot occur as long as the Polyakov loop stays small. Once the
Polyakov loop grows with increasing temperature, the chiral condensate
decreases, leading to the simultaneous crossovers. This handwaving
argument can be sophisticated in a more well-founded model. In the
Gocksch-Ogilvie model, which is derived from the lattice QCD in the
strong coupling and large dimensional expansion, it has been shown
that the chiral condensate is always finite if the Polyakov loop is
forced to be zero by hand~\cite{fuk03}.

\begin{figure}
\begin{center}
\includegraphics[width=5cm]{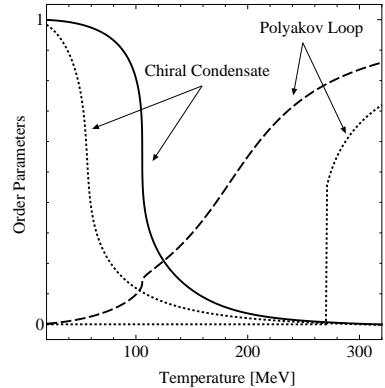}
\vspace{-5mm}
\caption{The chiral condensate and the Polyakov loop at
$\mu=321\,\text{MeV}$.}
\label{fig:order2}
\end{center}
\end{figure}

Fig.~\ref{fig:order2} is a prediction for the finite density case from
our model. The density is chosen at the chiral CEP, namely
$\mu=321\,\text{MeV}$ in this model. The temperature slope of the
chiral condensate diverges at $T=106\,\text{MeV}$. The Polyakov loop
has a longer tail because denser quark matter has more quarks breaking
the center symmetry. In this case it is hard to say that two
crossovers are simultaneous. In the future lattice simulation at
finite density, this prediction would be tested.

Although this model goes well for small $m_{\rm q}$, the level
repulsion discussed in Ref.~\cite{hat04} is not strong enough to lead
to the perfect locking between the chiral and deconfinement phase
transitions for all $m_{\rm q}$, as shown in Fig.~\ref{fig:phase}. It
is still an open question to clarify the full dynamics linking the
chiral and deconfinement phase transitions entirely.

The work involving Fig.~\ref{fig:phase} was performed in collaboration
with Yoshitaka Hatta.

\end{document}